\documentclass[aps]{revtex4}
\usepackage[colorlinks=true, pdfstartview=FitV, linkcolor=blue, citecolor=red, urlcolor=magenta, breaklinks=true]{hyperref}
\usepackage{amsfonts}
\usepackage{color,graphics}
\usepackage{graphicx}
\usepackage{amssymb}
\usepackage{amsmath}
\begin{document}
\title{Acoustic Black Holes and Universal Aspects of Area Products}
\author{M. A. Anacleto, F. A. Brito, E. Passos}
\email{fabrito, anacleto, passos@df.ufcg.edu.br}
\affiliation{Departamento de F\'{\i}sica, Universidade Federal de Campina Grande, 58109-970, Campina Grande, Para\'{\i}ba, Brazil}
\begin{abstract}
In this paper we derive acoustic metrics in the (3+1)-dimensional Abelian Higgs model with higher derivative terms. We have found acoustic metrics that are conformally related to the Reissner-Nordstr\"om and Kerr-Newman metrics. The universal aspects of area products which depend only on quantized quantities such as conserved electric charge and angular momentum are also addressed. We relate these areas with entanglement entropy of acoustic black holes in BEC systems. We also have shown there is an equivalence between microscopic descriptions of axisymmetric acoustic black hole entropy in a BEC system in four and two dimensions. Particularly, the system seems to develop an analogue of the Kerr/CFT correspondence.

\end{abstract}
\maketitle
\pretolerance10000
\section{Introduction}
Acoustic black hole metric has been extensively explored by several authors in different physical situations~\cite{Volovik,Unruh,Cadoni,MV}. 
The connection between black hole physics and
the theory of supersonic acoustic flow was established in 1981 by Unruh~\cite{Unruh} and pointed out that one can observe Hawking radiation of acoustic black hole in laboratory. On the other hand, from the viewpoint of observation, it is quite difficult to detect Hawking radiation of gravitational black holes because the Hawking temperature is quite low compared with cosmic microwave background (CMB) temperature.
Thus, the analog models have been developed to investigate the Hawking radiation and other phenomena for understanding quantum gravity on black holes in the laboratory.
Many fluid systems have been investigated on a variety of analog models of acoustic black holes, including gravity wave~\cite{RS}, water~\cite{Mathis}, slow light~\cite{UL}, optical fiber~\cite{Philbin} and electromagnetic waveguide~\cite{RSch}. The models of superfluid helium II~\cite{Novello}, atomic Bose-Einstein condensates (BEC)~\cite{Garay} and one-dimensional Fermi degenerate noninteracting gas~\cite{SG} have been proposed to create an acoustic black hole geometry in the laboratory.

Until very recently no analog formulae to the Hawking-Bekenstein entropy were known for acoustic black holes. However, in Ref.~\cite{rinaldi} was shown that such analogs indeed show up in a BEC system as one considers {\sl entanglement entropy} associated to phonons created via the Hawking mechanism --- see also \cite{HuiHui,Anacleto:2014apa}. One should stress that this entropy which depends on the horizon area of the acoustic black holes has nothing to do with the thermodynamic entropy of the fluid which is zero for a BEC system. Since the entanglement entropy of acoustic black holes is also given in terms of horizon area, thus it seems to be very natural to look for universal aspects of the horizon area products in systems other than gravitational ones. This new analog related to entanglement entropy comes to join the other well-known analogs mentioned above in view of providing a better understanding of gravitational systems. For logarithmic corrections, for instance, see \cite{KM-2000}. In the following we are going to show several effects whose resemblances with those in gravitational systems indeed happen. 

The purpose of this paper is to investigate the universal aspects of area products in the relativistic version of acoustic black holes derived from the Abelian Higgs model including terms of higher derivatives. It is worth mentioning that higher derivative terms have been introduced in theories of gravity with the aim of obtaining a consistent quantum theory of gravity~\cite{Uzan}.
In our model the contribution of higher derivative terms in the Lagrangian generates terms of fourth order for the fluid velocity field for the acoustic black hole metric. 
This is one of the generalizations of Schwarzschild acoustic black holes in four-dimensions by introducing analogs of gauge charge and/or angular momentum.  As we shall see, this will correspond to introducing a term of charge for  Reissner-Nordstr\"om (RN) metric. Plus, we will also show that the acoustic metric is conformally related to the  RN metric. 
For many purposes, since surface gravity and Hawking temperature are conformal invariants this is sufficient for analyzing basic features of the Hawking radiation process. On the other hand, if we are interested in analyzing the region close to an event horizon, the conformal factor can be neglected. Finally, we also find a sector in the acoustic black hole metric obtained in our model that will be identified with an acoustic Kerr-Newman (KN) black hole.

A relativistic version of acoustic black holes from the Abelian Higgs model has been presented in \cite{Xian}. More
recently, the metric of an acoustic black hole was obtained from a model of relativistic noncommutative fluid and also from a model with a Lorentz violating background fluid~\cite{ABP}. 
Differently of the most cases studied, we consider the acoustic black hole metrics obtained from a relativistic fluid model with higher derivatives terms. 
The effects of this set up are such that the fluctuations of the fluids are also affected. As a consequence the Hawking temperature is directly affected by terms of higher derivatives. Namely, one can now obtain analogs of RN black holes which have zero Hawking temperature for the extremal case. And more important, 
the universal aspects of area products which depend only on quantized quantifies such as conserved electric charge and angular momentum are now possible to be computed, since higher derivative terms lead to more than one horizon. Thus, we are able to establish a relationship between these area products due to Cauchy (inner) ($-$) and event (outer) ($+$) horizons as follows \cite{Visser}
 \begin{equation}\label{area-products}
A_+^2\geq A_+ A_-
\end{equation}
and  associate them with microscopic (entanglement) entropy of acoustic black holes in BEC systems. For instance, in Ref.~\cite{rinaldi} was shown that the entropy of an (1+1) - dimensional acoustic black hole in Bose-Einstein condensates,  in dilute BEC gas in a box of size $L$, in the hydrodynamic limit,  only depends on the size of the box in the near-horizon regime, as for (1+1) - dimensional gravitational black holes. Furthermore, the entropy found in such systems can be also expressed as
\begin{equation}\label{bh-bec-entropy}
S_{BH}\sim \tilde{n}(\omega),
\end{equation}
where $\tilde{n}(\omega)$ is the number of modes populating  a region inside the box of size $L$ with certain frequency $\omega$.
Since the black hole entanglement entropy is given in terms of the outer horizon area $S_{BH}\sim A_+\geq \sqrt{A_+A_-}$,  then as we shall see later, this relation between equations (\ref{area-products}) and (\ref{bh-bec-entropy}) suggests the area products for acoustic black holes should be quantized as in most of gravitational systems \cite{Visser}. Furthermore, in the latter case it is already well known that the microscopic description of the entropy of four and five-dimensional gravitational black holes can be computed in terms of string theory which can be understood as a (1+1) - dimensional conformal field theory \cite{cvetic-youm-vafa} --- {see also Kerr/CFT correspondence \cite{guica}}. As we shall see, in our present study we have found an equivalence that has a very high resemblance with this gravitational duality (one of the first clues of gauge/gravity duality, holography or Maldacena's correspondence \cite{ads-cft}).

\section{The Model}
\label{II}
The Lagrangian of the Abelian Higgs model including higher derivative gauge invariant terms we shall consider is the following
\begin{eqnarray}\label{our-model}
{\cal L}&=&-\frac{1}{4}F^2
+(D_{\mu}{\phi})^{\dagger} D^{\mu}{\phi}+ M^2{\phi}^{\dagger}{\phi}
+\frac{1}{\Lambda_{0}^2}(D_{\mu}D^{\mu}{\phi})^{\dagger} (D_{\nu}D^{\nu}{\phi})-b|{\phi}|^4,
\end{eqnarray}
where we have $F^2={F}_{\mu\nu}{F}^{\mu\nu}$, $F_{\mu\nu}=\partial_{\mu}A_{\nu}-\partial_{\nu}A_{\mu}$, $D_{\mu}\phi=\partial_{\mu}\phi-ieA_{\mu}\phi$ and $\Lambda_{0}$ is a
parameter with dimension of mass. The scalar field describes the superfluid density such a way that $\int{|\phi|^2\ d^3x}=N$ is the total number of atoms (in the nonrelativistic limit). 

{The higher derivative term in the Lagrangian corresponds to dispersion relation of phonons in atomic Bose-Einstein condensates. This dispersion relation is analogous to those previously considered in acoustic black holes \cite{SU} to address several issues, such as ultrashort-distance physics.} 

Now, in order to find the acoustic black hole metric, let us use the decomposition $\phi=\sqrt{\rho}e^{iS}$ in the original Lagrangian and neglect terms of order $ {\cal O}(\Box\sqrt{\rho}) $ (that can be negligible in the hydrodynamic region), we obtain \cite{Xian,ABP}
\begin{eqnarray}
\label{Lag2}
{\cal L}&=&-\frac{F^2}{4}+\left(u^2 + M^2\right)\rho-b\rho^2+\frac{\rho}{\Lambda_{0}^{2}}
\left[(\Box S)^2+(\partial_{\mu}S)^4+e^2A^2(2u^2-e^2A^2)
-4eA^{\mu}u^{\nu}\partial_{\mu}S\partial_{\nu}S \right],
\end{eqnarray}
 where $ u_{\mu}=\partial_{\mu}S-eA_{\mu}$, $ A^2=A_{\mu}A^{\mu}$ and $\Box=\partial_{\mu}\partial^{\mu}$ is the D'Alembert operator in Minkowski space. { As we shall discuss latter, in a superfluid (or BEC system) we have a quantum vortex as a hole with the superfluid circulating around the vortex with the phase difference around the vortex given by $\Delta S=2n\pi$}.

By using the Lagrangian (\ref{Lag2}) 
and linearizing the equations of motion around the background
$(\rho_{0}, S_{0})$, with $\rho=\rho_{0}+ \rho_{1}$ and {$S=S_{0}+\psi$}, whereas the vector field $ A_{\mu} $ remains unchanged, we find the equation of motion for a linear acoustic disturbance ψ given by a Klein-Gordon equation in a curved space with a higher derivative source~\cite{SU}
\begin{equation}
\label{eqkg}
\frac{1}{\sqrt{-g}}\partial_{\mu}\sqrt{-g}g^{\mu\nu}\partial_{\nu}\psi={\frac{1}{\Lambda_{0}^2}\partial_{\mu}(\rho_{0}\partial^{\mu}\Box\psi)},
\end{equation}
where the index $\mu=t,x_{i} $, $ i=1,2,3$.   In the following we shall define $c^2_{s}=b\rho_{0}/2w^2_{0}$ (the local sound speed in the fluid),  {$v^{i}=v^i_{0}/w_{0}$} (velocity of the flow),  $\vec{v}_{0}=\nabla S_{0}+e\vec{A}$,
\begin{equation}\label{L0-L}
w_{0}=-\dot{S}_{0}+eA_{t},\qquad\mbox{and}\qquad \Lambda^2=\Lambda^{2}_{0}/w^{2}_{0}.
\end{equation}
In terms of the inverse of $g^{\mu\nu}$ we have the metric of an acoustic black hole 
$g^{eff}_{\mu\nu}=\frac{b\rho_{0}}{2c_{s}\sqrt{f}}g_{\mu\nu}$,
where neglecting terms of order $ {\cal O}(1/\Lambda^4) $, we find
\begin{eqnarray}\label{metric-v4}
{g}_{00}&=&-\left(c^{2}_{s}-v^2\right) 
-\frac{\left(2c^{2}_{s}-4v^2+4v^4-6c^2_{s}v^2 \right)}{\Lambda^2},
\nonumber\\
g_{0i}&=&g_{i0}=-v^{i}-\frac{4}{\Lambda^2}\left(1+c^{2}_{s}-v^2\right)v^{i} ,
\nonumber\\
g_{ij}&=&\left( 1+\frac{4}{\Lambda^2} \right)\left(1+c^{2}_{s}-v^2\right)\delta^{ij} 
+v^{i}v^{j}+\frac{2c^{2}_{s}}{\Lambda^2}\left(1-v^2\right)\delta^{ij} ,
\nonumber\\
f&=&1+c^{2}_{s}-v^2+O(1/\Lambda^2).
\end{eqnarray}
This metric simply depends on the density $\rho_0$, the local sound speed in the fluid $c_s$ and the velocity of flow $\vec{v}$. 
In the limit $ \Lambda^2 \longrightarrow \infty $ the result in \cite{Xian} is recovered.

By using the Lagrangian (\ref{Lag2}) we can find the following continuity equation 
\begin{equation}
\label{cont2}
\partial_{\mu}\left[\rho u^{\mu}+\frac{2\rho}{\Lambda_{0}^{2}}u^{\mu}u^{\nu}u_{\nu}+\frac{\rho}{\Lambda_{0}^{2}} \partial^{\mu}\partial_{\nu}u^{\nu} \right]=0.
\end{equation}
Keeping $\rho\equiv\rho_0$ as a constant density, assuming static fields and disregarding the second term in comparison with the first one for sufficiently large $\Lambda_0$ we find 
\begin{equation}
\label{cont3}
-\nabla^2\Psi(r)=\Lambda_0^2\Psi(r),\qquad \mbox{where $\qquad\nabla\cdot \vec{v}(r)=\Psi(r)$, $\qquad$with$\qquad$ $v^i=u^i,\:\: i=1,2,3$}.
\end{equation}
We shall consider spherically symmetric solutions. Notice this is precisely a radial Schroedinger-like equation for a free particle. Once we obtain $\Psi(r)$ we can find the radial velocity $v_r(r)$. The normalized and regular solution at $r=0$ is given by
\begin{equation}
\Psi(r)=\sqrt{\frac{\Lambda_0}{\pi}}\frac{\sin{(\Lambda_0 r)}}{r}.
\end{equation}
Thus, the radial velocity can be readily given by integrating the equation $\nabla\cdot \vec{v}(r)=\Psi(r)$ to find
\begin{eqnarray}\label{v-radial0}
v_r(r)&=&\sqrt{\frac{\Lambda_0}{\pi}}\frac{\sin{\Lambda_0r-\Lambda_0r\cos{\Lambda_0r}}}{\Lambda_0^2r^2}\nonumber\\
&\propto&{{\frac{1}{\sqrt{r_h}}}}, \qquad \mbox{with $\qquad r\simeq r_h=2\pi/\Lambda_0$}
\end{eqnarray}
where we have assumed the near horizon condition $r=r_h+\epsilon$ ($\epsilon\to0$) which also implies
\begin{equation}\label{v-radial}
v_r(r)\propto{\frac{1}{\sqrt{r-\epsilon}}}\simeq\frac{1}{\sqrt{r}}\left(1+\frac12\frac{\epsilon}{r}\right)\simeq \frac{1}{\sqrt{r}}+{\cal O}(\epsilon/r).
\end{equation}
{Thus, from Eq.~(\ref{v-radial0}) and leading term of (\ref{v-radial}) we find the relationship 
\begin{eqnarray}\label{v-radial-2}
v_r(r)^2&\propto&\Lambda_0(r)\nonumber\\
&\propto&{{\frac{1}{{r}}}},\nonumber\\ 
v_r(r)^2&=&2U(r), \qquad \mbox{with \qquad $U(r)=\frac{M}{r}$}.
\end{eqnarray}
Notice that in the last step we have identified a gravitational-like potential $U(r)=M/r$ due to an object with `mass' $M$.
 }
 
{Furthermore, from  Eq.~(\ref{L0-L}), $\Lambda^2=\Lambda_0^2/w_0^2$ for static field configurations $w_0=eA_t=e/r$ and r.h.s. of  Eqs.~(\ref{v-radial-2}), $\Lambda_0={2M}/{r}$, we can identify a squared `screened charge' $e^2=4M^2/\Lambda^2$. Alternatively, from the first Eq.~(\ref{metric-v4}) the fourth-order contribution gives
\begin{eqnarray}\label{eq-Q-M}
\frac{1}{\Lambda^2}{4v_r^4(r)}&=&\frac{1}{\Lambda^2}\frac{16M^2}{r^2}\nonumber\\
&=&\frac{1}{\Lambda^2}\frac{Q^2}{r^2}\nonumber\\
&\equiv& \frac{4e^2}{r^2}, \qquad\mbox{with \qquad} \frac{Q^2}{\Lambda^2}=4e^2.
\end{eqnarray}
As we shall properly address later, the `screened charge' $Q/\Lambda$ will be the right electric charge of the charged black hole. The `bare charge' $Q$ that easily satisfies Eq.~(\ref{eq-Q-M}) for $Q^2=16M^2$ will gain another interpretation. 
}
\section{Acoustic Reissner-Nordstr\"om metric}

Now  we consider the speed of sound $c_{s} $ to be a
position independent constant. 
{So using 
(\ref{v-radial-2}) for radial velocity we can assume}
$ v_{r}=\sqrt{2M/r}$
and we obtain a RN metric type 
$ ds_{RN}^2\propto\frac{\rho_0}{\sqrt{f}} ds^2$, where 
\begin{eqnarray}
ds^2=-F_{eff}(v_{r})d\tau^{2}+\frac{fc^{2}_{s}dr^{2}}{F_{eff}(v_{r})}+r^{2}d\Omega^2,
\end{eqnarray}
{with
\begin{eqnarray}
F_{eff}(r)&=&\left(1+\frac{2}{\Lambda^2}\right)c^{2}_{s}-\frac{2M_{eff}}{r}
+\frac{1}{\Lambda^2}\frac{Q^{^2}}{r^{2}},
\end{eqnarray}
where } $M_{eff} =M\left(1+\frac{4+6c^{2}_{s}}{\Lambda^2}\right) $, $M$ is the mass and  $ Q^{2}=16M^2$ is identified with {the `bare' electric charge.} 
This defines a `particle' with a charge-to-mass ratio $Q/M=4$ similar 
to what happens to electron, for instance,  whose charge exceeds its mass. In the limit $ \Lambda^2\longrightarrow \infty $ we have a Schwarzschild metric type. 
The result is conformal to the RN metric type.
If we consider a region near the event horizon, the conformal factor can be simply taken to be a constant and can be 
neglected~\cite{MV, Cadoni}.
The horizon is given by the coordinate singularity $g_{rr}=0$, that is
\begin{eqnarray}
r_{\pm}=\frac{\lambda}{2}\left[\tilde{r}_{s} \pm\sqrt{\tilde{r}^{2}_{s}-4\tilde{r}^{2}_{Q}}\right],
\end{eqnarray}
{where $ \lambda=\left(1+{2}/{\Lambda^2}\right)^{-1} c^{-2}_{s}$, $\tilde{r}_{s}=2M_{eff}$ and 
$ \tilde{r}^{2}_{Q}=Q^2_{eff}\equiv{Q^2}/{\lambda\Lambda^{2}}=4e^2/\lambda$.} The condition 
$\tilde{r}^2_{s}=4\tilde{r}^2_{Q}$
corresponds to an extremal RN black hole. The Hawking temperature is given by
\begin{equation}
T_{H}=\frac{F^{\prime}(r_{+})}{4\pi}=\frac{1}{4\pi}\left(\frac{\tilde{r}_{s}}{{r}^2_{+}}-\frac{2\tilde{r}^{2}_{Q}}{{r}^3_{+}}\right).
\end{equation}
Therefore, for $ \tilde{r}^2_{s}=4\tilde{r}^2_{Q}$, we obtain $T_{H}=0$ just as in the gravitational RN black holes. {Notice that the effective mass and charge are in fact the quantities that determine the RN acoustic black hole, since they enjoy the condition of existence of the black hole $M_{eff}\geq Q_{eff}$ ($\tilde{r}^2_{s}\geq4\tilde{r}^2_{Q}$). This is valid for finite $\Lambda$, with the non-charged (Schwarzschild) limit as $\Lambda$ becomes very large. On the other hand, the `bare' quantities $M$ and $Q$ cannot satisfy this relationship since $M<Q$, according to the charge-to-mass ratio imposed above. In this sense, only the effective mass and charge are physically acceptable to describe analogs of gravitational mass and electric charge of a RN black hole.}

\section{Acoustic Kerr-Newman metric}
The acoustic metric can also be written in a KN form in Boyer-Lindquist coordinates  with $v_{r}\neq 0$, $ v_{\phi}\neq 0 $ and $ v_{\theta}=0 $. In the nonrelativistic limit $ds_{KN}^2=\frac{b\rho_0}{2c_s\sqrt{f}} ds^2$, with
\begin{eqnarray}
\label{kn}
ds^2&=&-F(v)d\tau^2+\frac{fc^{2}_{s}}{F(v_r)}dr^2
-2\varpi\sin^{2}\theta d\varphi d\tau + r^{2}d^{2}\Omega.
\end{eqnarray}
Now setting $v_{r}=\sqrt{2M/r}$ and $ v_{\phi}=J/Mr=a/r $  into (\ref{kn}), 
we obtain, $ ds_{KN}^2\propto\frac{\rho_0}{\sqrt{f}} ds^2$
with the following changes $ F(v_{r})\Longrightarrow  F(r) $ and 
$ F(v)\Longrightarrow {\cal F}(r)=F(r)- \left(1+(4+6c^{2}_{s})/\Lambda^2\right)a^2/r^2$, being
\begin{eqnarray}
F(r)=\left(1+\frac{2}{\Lambda^2}\right)c^{2}_{s}-\frac{2M_{eff}}{r}
+\frac{1}{\Lambda^2}\frac{(Q^{^2}+a^2)}{r^{2}},
\end{eqnarray}
and the gyromagnetic ratio given by
\begin{eqnarray}
\varpi=\frac{a}{r}\left[1+\frac{1}{2\Lambda^2}\left(4+6c^{2}_{s}-\frac{Q^2+a^2}{Mr}-\frac{4a^2}{r^2}\right)\right],
\end{eqnarray}
where we have identified $Q^2+a^2= 16M^2$
and $a$ is the angular momentum per unit mass.  Again, this defines a `particle' similar 
to electron whose charge $Q$ and $a$ exceed its mass. The horizons are
\begin{eqnarray}
r_{\pm}=\lambda M_{eff} \pm
\lambda\sqrt{M_{eff}^{2}-Q^{2}_{eff}-a^{2}_{eff}},
\end{eqnarray}
where $Q^{2}_{eff}=Q^2/\lambda\Lambda^2$ and $a^{2}_{eff}=a^2/\lambda\Lambda^2$. The areas of these horizons are
\begin{eqnarray}
A_{\pm}=4\pi \left[r^{2}_{\pm}+\lambda^2 a^2_{eff} \right].
\end{eqnarray}
Then we have the event horizon area
\begin{eqnarray}
\label{A2}
A_{+}^2&=&(8\pi)^{2}\lambda^4\Big[\frac{Q^{4}_{eff}}{4}+M^{2}_{eff}a^{2}_{eff}+2 M^{4}_{eff}
-2M^{2}_{eff}(Q^{2}_{eff}+a^{2}_{eff})   
\nonumber\\
&+& M_{eff}(2M^{2}_{eff}-Q^{2}_{eff})(M_{eff}^{2}-Q^{2}_{eff}-a^{2}_{eff})^{1/2}\Big],
\end{eqnarray}
and the area product~\cite{Visser}
\begin{eqnarray}\label{area-product-2}
A_{+}A_{-}&=&(8\pi)^2\lambda^{4}\left[M^{2}_{eff}a^{2}_{eff}+\frac{Q^{4}_{eff}}{4}\right]
\nonumber\\
&=&\frac{(8\pi)^2}{c^{4}_{s}}\left[\frac{J^2}{c^{2}_{s}\Lambda^2}+\frac{Q^4}{4\Lambda^4}\right]\!\!,
\end{eqnarray}
where we have taken terms up to $1/\Lambda^4$ order in the last step. We should also recall that we are adopting {natural units $G=\hbar=c=1$.}
This product is independent of the {effective mass $M_{eff}$} of the acoustic black hole, and depends only on $ J $  and  $Q$. 
Thus, by comparison between equations (\ref{A2}) and (\ref{area-product-2}) we can easily find a relationship between the area products due to outer and inner/outer horizons, that is
\begin{equation}
A_+^2\geq A_+ A_-.
\end{equation}
For the extremal case, i.e., using $M^2_{eff}= Q^{2}_{eff}+a^{2}_{eff}$ into (\ref{A2}), it is easy to show that the inequality saturates, that is, $A_+^2= A_+ A_-$.

In the slow-rotation limit we again obtain the (3+1) KN metric, $ds_{KN}^2$,  of a rotating charged black hole. Neglecting terms of order $ O(\tilde{a}^2) $~\cite{Berti} we have that $ {\cal F}(r)=F(r) $ and the gyromagnetic ratio  is now given by
\begin{eqnarray}
\varpi=\frac{\tilde{a}M^2}{r}\left[1+\frac{2+3c^{2}_{s}}{\Lambda^2}\right]-\frac{\tilde{a}MQ^2}{2\Lambda^2 r^2} +O(\tilde{a}^3).
\end{eqnarray}
where, $ \tilde{a}\equiv a/M=J/M^2 $ (rotation parameter). 

For a condensate with density distribution with axial symmetry which implies axisymmetric acoustic black holes  we can deal with a quantized area product as follows. {Let us consider a condensate in the nonrelativistic limit of our Lagrangian (\ref{our-model}) for  $\Lambda\to\infty$, which leads to a Gross $-$ Pitaevskii like equation}. The velocity given in terms of the phase, i.e. $\vec{v}=\nabla S/M$ is not always irrotational,  because there are singularities in the phase $S$ in BEC systems. Thus, the line integral of $\vec{v}$ along a closed contour $C$ is
\begin{eqnarray}
\oint_C{\vec{v}\cdot d\vec{l}}=\frac{1}{M}\oint_C{\nabla S\cdot d\vec{l}}=\frac{\Delta S}{M}=\frac{2\pi \ell}{M},
\end{eqnarray}
which is quantized. This simply recover our previously considered velocity $v_\phi=J/Mr=a/r$, with quantized circulation $a=\ell/M$. Now applying the area product (\ref{area-product-2}) we find
\begin{eqnarray}\label{area-product-3}
A_{+}A_{-}=\frac{(8\pi)^2}{c^{4}_{s}}\left[\frac{\ell^2}{c^{2}_{s}\Lambda^2}+\frac{
Q^4}{4\Lambda^4}\right]\!\!, \qquad \ell=1,2,...,
\end{eqnarray}
which is written in terms of quantized angular momentum $J=\ell$. In general, our system have both angular momentum $J$ and charge $Q$ that should  also be quantized as $Q=n\, e$, where $e$ is the elementary electric charge. However, in the aforementioned BEC system with large $\Lambda$, let us take only the leading order term of (\ref{area-product-3}), i.e., 
\begin{eqnarray}
A_{+}A_{-}=(8\pi)^2({\ell^2}/{c^{6}_{s}\Lambda^2})
\end{eqnarray}
According to extremal limit, that we previously mentioned, we should find
\begin{eqnarray}
A_+^2=(8\pi)^2({\ell^2}/{c^{6}_{s}\Lambda^2})
\end{eqnarray}
Thus, the (3+1) - dimensional axisymmetric acoustic black hole entropy in a BEC system is
\begin{eqnarray}\label{ours-entropy}
S_{BH}^{3+1}=\frac14A_+=2\pi({\ell}/{c^{3}_{s}\Lambda}).
\end{eqnarray}
As we anticipated in the introduction the (1+1) - dimensional acoustic black hole entropy in BEC system is found to be 
\begin{eqnarray}\label{entropy-rinaldi}
S_{BH}^{1+1}=2\pi\left(\tilde{n}(\omega)/\tilde{\Lambda}\right)
\end{eqnarray}
where $\tilde{\Lambda}=(6\omega/\pi T_H)$ and $\tilde{n}(\omega)$ is the number of modes populating a region inside a one-dimensional box of size $L$ with certain frequency $\omega$. The number of modes $\tilde{n}$ on a line with odd (or even) number of nodes completing entire cycles plays the same role of doing complete circulations around the core of the vortices in a Bose-Einstein condensate given by $\ell$. Thus, up to simple identification of scale ${\Lambda}$ with $\tilde{\Lambda}$, we have here two {\it completely equivalent microscopic descriptions  of axisymmetric acoustic black hole entropy in a BEC system in four and two dimensions}. This observation has a very high resemblance with the microscopic description of the entropy of four and five-dimensional gravitational black holes in terms of string theory which can be understood as a (1+1) - dimensional conformal field theory \cite{cvetic-youm-vafa}. 

{In order to support this equivalence, let us now focus on the Kerr/CFT correspondence \cite{guica}. In this correspondence it is shown that quantum gravity on NHEK 
(near-horizon extreme Kerr) is holographically dual to a 2d conformal field theory with central charge $c_L=12J$. The microscopic entropy of the 4d Kerr black hole with angular momentum $J$ is then given in terms of the Cardy's formula 
\begin{equation}\label{cardy}
S_{CFT}^{1+1}=\frac{\pi^2}{3}\,c_L\,{T}_{L},
\end{equation}
where $T_L=1/2\pi$ is the left-moving dimensionless temperature. As was shown in \cite{guica}, even though extreme Kerr has zero Hawking temperature, the quantum fields outside the horizon are not in a pure state. Plugging this temperature and the central charge into Eq.~(\ref{cardy}) reveals that the microscopic description of the Kerr black hole entropy is 
\begin{equation}
S^{1+1}_{CFT}={2\pi J}=\frac{A}{4}=S^{3+1}_{macro},
\end{equation}
which precisely reproduces the macroscopic Bekenstein-Hawking entropy of the Kerr black hole.

Now returning to our system, note that we can rewrite Eq.~(\ref{entropy-rinaldi}) as follows
\begin{eqnarray}\label{entropy-rinaldi-2}
S_{BH}^{1+1}=\frac{\pi^2}{3}\,c\,\check{T}_{H},\qquad c={\tilde{n}}, \qquad \check{T}_H=\frac{T_H}{\omega},
\end{eqnarray}
that is analogous to the Cardy's formula (\ref{cardy}) with $\tilde{n}$ playing the role of a `central charge' and  $\check{T}_H$ being analogous to the left-moving dimensionless temperature $T_L$. Now choosing $\check{T}_H=1/2\pi$ and $c=12\ell/\Lambda c_s^3$ we achieve an equivalence analogous to the Kerr/CFT correspondence, 
that is, the microscopic description of the 2d entropy
\begin{equation}
 S_{BH}^{1+1}=2\pi (\ell/\Lambda c_s^3)=\frac14A_+=S_{BH}^{3+1},
\end{equation}
again precisely reproduces the macroscopic Bekenstein-Hawking entropy (\ref{ours-entropy}) of the analogous Kerr black hole.
}
{This analogous Kerr/CFT correspondence present  in our system seems to be expected since the 2d entropy (\ref{entropy-rinaldi}) found in \cite{rinaldi} in the hydrodynamic limit is the same as the one predicted by conformal field theory via brick wall model for gravitational black holes.  Furthermore, the 4d entropy (\ref{ours-entropy}) is in full accord with entropy of 4d extremal Kerr black holes.}

\section{Conclusions}
\label{conclu}
In this paper we have considered the extended Abelian Higgs model by introducing higher derivative terms. The acoustic black hole metric in our model can be identified with acoustic Reissner-Nordstr\"om and  Kerr-Newman black holes. We also have found general aspects of area products in terms of outer and inner horizons and have no dependence with the black hole mass. They depend only on quantized charge and angular momentum. Finally, and more interesting, we also observed that the microscopic entropy in the (3+1) - dimensional axisymmetric acoustic black hole in a BEC system can also be described by the microscopic entropy in the (1+1) - dimensional acoustic black hole in a BEC system. Thus, we have found an equivalence that has a very high resemblance with a correspondence between microscopic description of gravitational black holes entropy in five/four dimensions and two-dimensional conformal field theory \cite{cvetic-youm-vafa} (one of the first clues of the gauge/gravity duality, holography or Maldacena's correspondence \cite{ads-cft}) --- see also the recent fluid/gravity duality \cite{Bredberg:2011jq}. {Particularly, our system seems to develop an analogue of the Kerr/CFT correspondence established in \cite{guica}.} {Although we have not addressed more properties to support (or proof) this equivalence, we find that we have presented strong enough evidences. However, further investigations should be done elsewhere, for instance by considering other solutions in four or arbitrary dimensions.}
\acknowledgments
We would like to thank Mirjam Cvetic for invaluable discussions. The authors also thank CNPq, CAPES, PNPD/PROCAD-CAPES for partial financial support.

\end{document}